\documentclass[twocolumn,prl,aps,epsf,showpacs,floatfix]{revtex4}
 
\usepackage{graphicx}
\usepackage{amssymb}
\usepackage[dvips]{color}

\begin{document}

%\begin{frontmatter}

\title{
Improved measurement of the $\pi \rightarrow \mbox{e} \nu$ branching ratio
}

\author{A.Aguilar-Arevalo$^1$,
M. Aoki$^2$,
M. Blecher$^3$,
D.I. Britton$^4$,
D.A. Bryman$^5$,
D. vom Bruch$^5$,
S. Chen$^6$,
J. Comfort$^7$,
M. Ding$^6$,
L. Doria$^8$,
S. Cuen-Rochin$^5$,
P. Gumplinger$^8$,
A. Hussein$^9$,
Y. Igarashi$^a$,
S. Ito$^2$,
S.H. Kettell$^b$,
L. Kurchaninov$^8$,
L.S. Littenberg$^b$,
C. Malbrunot$^{5,*}$,
R.E. Mischke$^8$,
T. Numao$^8$,
D. Protopopescu$^4$,
A. Sher$^8$,
T. Sullivan$^5$,
D. Vavilov$^8$,
K. Yamada$^2$\\
(PIENU Collaboration)}

\affiliation
{$^1$Instituto de Ciencias Nucleares, Universidad Nacional Aut\'onoma de Mexico, D.F. 04510 M\'exico\\
$^2$Graduate School of Science, Osaka University, Toyonaka, Osaka 560-0043, Japan\\
$^3$Physics Department, Virginia Tech., Blacksburg, VA 24061, USA\\
$^4$Physics Department, University of Glasgow, Glasgow, UK\\
$^5$Department of Physics and Astronomy, University of British Columbia,
Vancouver, B.C. V6T 1Z1, Canada\\
$^6$Department of Engineering Physics, Tsinghua University, Beijing, 100084, China\\
$^7$Physics Department, Arizona State University, Tempe, AZ 85287, USA\\
$^8$TRIUMF, 4004 Wesbrook Mall, Vancouver, B.C. V6T 2A3, Canada\\
$^9$University of Northern British Columbia, Prince George, B.C. V2N 4Z9, Canada\\
$^a$KEK, 1-1 Oho, Tsukuba-shi, Ibaraki 305-0801, Japan\\
$^b$Brookhaven National Laboratory, Upton, NY 11973-5000, USA\\
$^*$Present address: CERN, 1211 Geneva 21, Switzerland and Stefan-Meyer-Institut f\"ur subatomare Physik,
 Austrian Academy of Sciences, Boltzmanngasse 3, A-1090 Vienna, Austria
}

\date{\today}

\begin{abstract}

A new measurement of the branching ratio,
$R_{e/\mu} =\Gamma (\pi^+ \rightarrow \mbox{e}^+ \nu + \pi^+ \rightarrow \mbox{e}^+ \nu \gamma)/
\Gamma (\pi^+ \rightarrow \mu^+ \nu + \pi^+ \rightarrow \mu^+ \nu \gamma)$,
resulted in  $R_{e/\mu}^{exp} = (1.2344 \pm 0.0023 (stat) \pm 0.0019 (syst)) \times 10^{-4}$. 
This is in agreement with the standard model prediction and improves the test of
 electron-muon universality to the level of 0.1 \%.
\end{abstract}


\maketitle

%\section{Introduction}

The standard model (SM) assumes equal electro-weak couplings of 
the three lepton generations, a hypothesis known as lepton universality 
which is studied in high precision measurements of $\pi, K, \tau, B,$ and $W$ decays. 
A recent measurement of  $B^+ \rightarrow K^+ l^+ l^-$ decays \cite{lhcb},
where $l$ represents e or $\mu$,
 hinted at a possible violation 
of e-$\mu$ universality in second order weak interactions that
involve neutral and charged currents.
The branching ratio of pion decays,
$R_{e/\mu} =\Gamma (\pi \rightarrow \mbox{e} \nu (\gamma))/
 \Gamma (\pi \rightarrow \mu \nu (\gamma))$, where $(\gamma)$ indicates 
inclusion of associated radiative decays,
has been calculated in the SM with extraordinary precision 
to be $R_{e/\mu}^{SM}=(1.2352 \pm 0.0002) \times 10^{-4}$ \cite{vc,bmty}.
Comparison with the  latest experimental values,
 $R_{e/\mu}^{exp}=(1.2265 \pm 0.0034(stat) \pm 0.0044(syst)) \times 10^{-4}$
 \cite{triumf}
and $R_{e/\mu}^{exp}=(1.2346 \pm 0.0035(stat) \pm 0.0036(syst)) \times 10^{-4}$ 
\cite{psi},
 has provided one of the best tests of e-$\mu$ universality
 in weak interactions for the charged current, at the 0.2 \% level
 giving sensitivity to new physics beyond the SM up to
mass scales of $O$(500) TeV\cite{bmty}.
Examples of new physics
 probed include R-parity violating SUSY \cite{mjrm},
extra leptons \cite{endo} and leptoquarks \cite{lepto}.
In this paper, we present the first results from the PIENU experiment,
which improve on the precision of  $R_{e/\mu}^{exp}$ and the test of 
e-$\mu$ universality.

%\section{Experiment}

\begin{figure}[htb]
\centering
\includegraphics*[width=8.3cm]{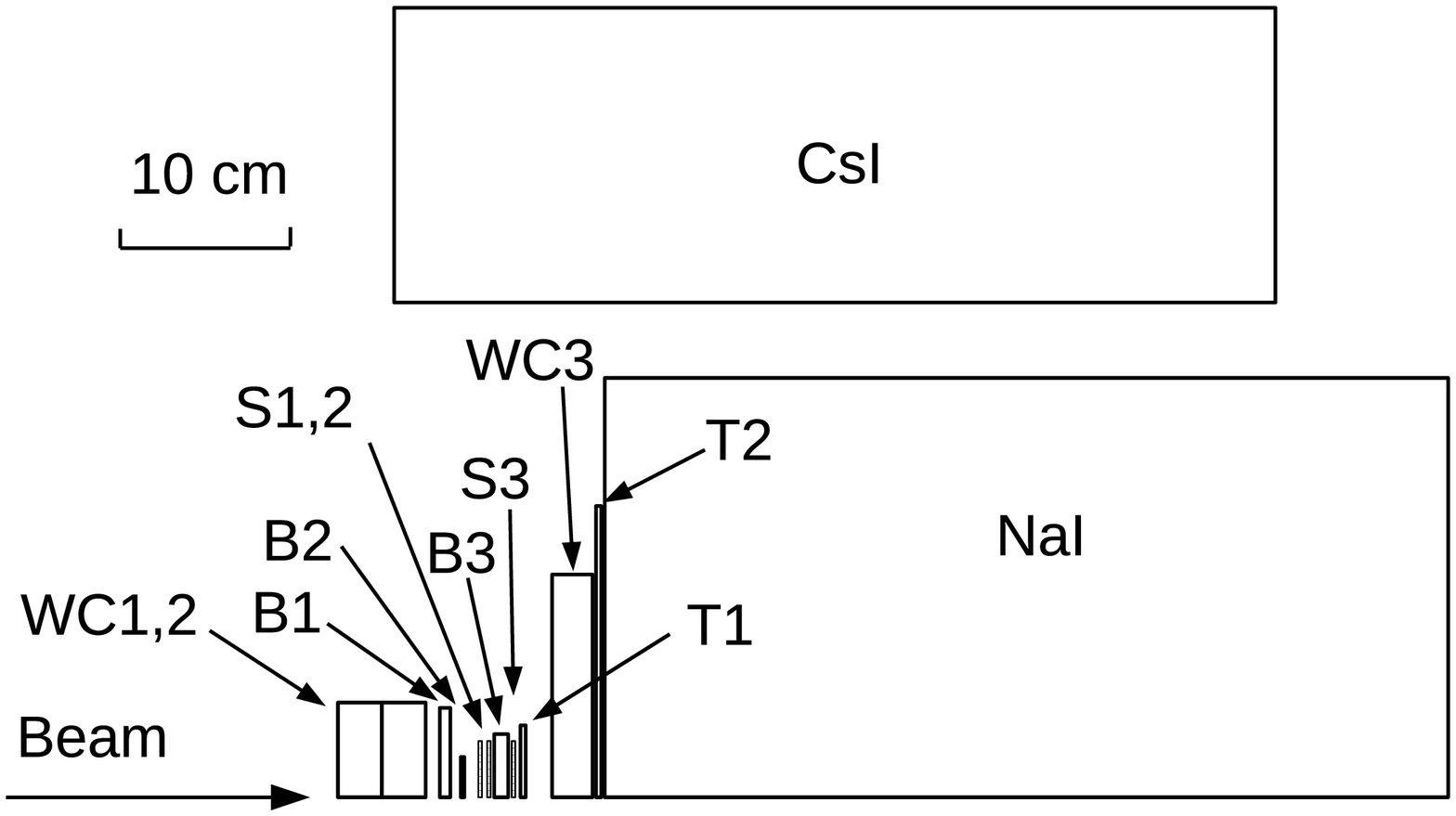}
\caption{Top half cross-section of the PIENU detector.
The cylindrical NaI(T$\ell$) crystal is surrounded by a
cylindrical array of CsI crystals as described in the text.}
\label{detector}
\end{figure}

The branching ratio 
$R_{e/\mu}$
is obtained from the ratio of positron yields from the
$\pi^+ \rightarrow \mbox{e}^+ \nu (\gamma)$ decay (total positron energy
E$_{e^+}=69.8$ MeV) 
and the $\pi^+ \rightarrow \mu^+ \nu (\gamma)$ decay followed by the 
$\mu^+ \rightarrow \mbox{e}^+ \nu \overline{\nu} (\gamma)$ decay
($\pi^+ \rightarrow \mu^+ \rightarrow \mbox{e}^+$, E$_{e^+}=0.5-52.8$ MeV) 
using pions at rest.
Figure \ref{detector} shows a schematic view of
the apparatus \cite{nimpaper} in which
a 75-MeV/$c$ $\pi^+$ beam from the TRIUMF M13 channel \cite{m13}
was degraded by two thin plastic scintillators
B1 and B2
and stopped in an 8-mm thick scintillator target
(B3) at a rate of $5 \times 10^4$ $\pi^+$/s.
Pion tracking  was provided by wire chambers (WC1 and WC2)
 at the exit of the beam line 
and two ($x$,$y$) sets of single-sided 0.3-mm thick planes
of silicon strip detectors, 
S1 and S2, located immediately
upstream of B3.

The positron calorimeter, 19 radiation lengths (r.l.) thick,
placed on the beam axis consisted of 
a 48-cm (dia.) $\times$ 48-cm (length) single-crystal
NaI(T$\ell$) detector \cite{bnl}
preceded by two thin plastic scintillators
(T1 and T2).
Two concentric layers of pure CsI crystals \cite{e949}
(9 r.l. radially, 97 crystals total) surrounded the NaI(T$\ell$) 
crystal to capture electromagnetic showers.
 Positron tracking was done by
an  ($x,y$) pair of Si-strip detectors (S3) and 
wire chambers (WC3) in front of the NaI(T$\ell$) crystal.

A positron signal, defined by a T1 and T2 coincidence,
occurring in a time window  --300 to 540 ns with respect to the incoming pion
 was the basis of the main trigger logic.
This was prescaled by a factor of 16 to form an unbiased trigger 
(Prescaled-trigger).
Events in an early time window  6 to 46 ns
and events with $E_{e^+} > 46$ MeV in the calorimeter
 provided other triggers 
(Early- and HE-triggers), which
included most $\pi^+ \rightarrow \mbox{e}^+ \nu$ decays.
The typical trigger rate (including monitor triggers) was  600 Hz.

\begin{figure}[htb]
\centering
\includegraphics*[width=8.3cm]{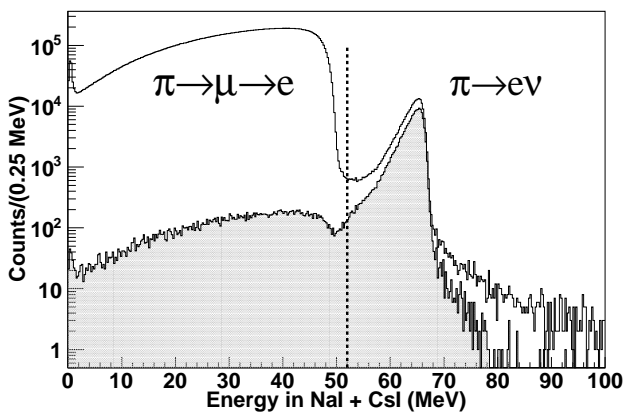}
\caption{
Energy spectra of positrons in the time region 5 to 35 ns
without and with (shaded) background-suppression cuts
(see the text).
The vertical line at 52 MeV indicates the E$_{cut}$ position.
}
\label{energy}
\end{figure}

%\section{Analysis}

%\subsection{Event selection}

Events originating from stopped pions were selected 
based on their energy losses
in B1 and B2. 
Any events with extra activity in the beam 
and positron counters (B1, B2, T1 and T2)
in the time region of --7 to 1.5 $\mu$s with respect to the
pion stop were rejected.
About 40 \% of events survived the cuts.
A fiducial cut for positrons entering the NaI(T$\ell$) detector
required a track at WC3 to be within 60 mm of the beam axis
to reduce electromagnetic shower leakage from the crystal.

\begin{figure}[thb]
\centering
\hspace*{0.3cm}
\includegraphics*[width=8.3cm]{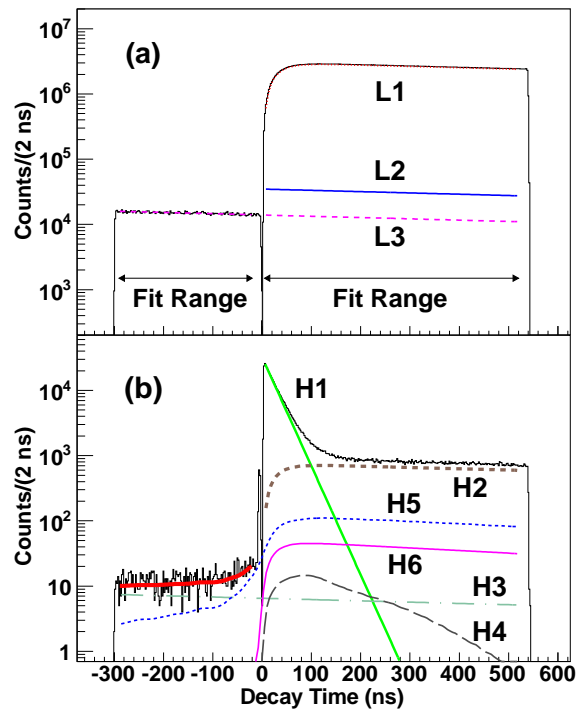}
\caption{(Color online)
Time spectra of positrons (thin line histograms) 
in the (a) low- and (b) high-energy regions separated at E$_{cut}$. 
The notches at $t =0$ ns are due to a veto for prompt pion decays,
and the peak at --3 ns in (b) is due to positrons in the beam.
Each curve, labeled with the corresponding component described in the text,
indicates the amplitude in the fit. L1 and part of L3 significantly
overlap with the data.
The thick solid line in (b) for $t<0$ ns shows the fit.
The fit for the other regions is almost indistinguishable from the data
and is omitted here.}
\label{time}
\end{figure}

The summed NaI(T$\ell$) and CsI energy for
positrons in the time region 5 to 35 ns
is shown in Fig. \ref{energy}.
The time spectra for events
in the low- and high-energy regions 
separated at E$_{cut}=52$ MeV are shown in Fig. \ref{time}.
Events satisfying the Early-trigger or Prescaled-trigger 
filled the low-energy histogram (Fig. \ref{time}a) 
and HE-trigger events filled the high-energy histogram (Fig. \ref{time}b). 
There were $4 \times 10^5$ $\pi^+ \rightarrow {\rm e}^+ \nu$ events at this stage.
The raw branching ratio was determined
from the simultaneous fit of these timing distributions.
To reduce possible bias, 
the raw branching ratio was shifted (``blinded'')
by a hidden random  value within 1 \%.
Prior to unblinding,
all cuts and corrections were determined and the stability of the result
against variations of each cut
was reflected in the systematic uncertainty estimate.

%\subsection{Fitting the time spectra}

%%\subsubsection{Low-energy time spectrum}

In the low-energy time spectrum,
the main components
were $\pi^+ \rightarrow \mu^+ \rightarrow \mbox{e}^+$ decays at rest (L1), 
$\mu^+ \rightarrow \mbox{e}^+ \nu \overline{\nu}$ 
decays (L2, about 1 \% of L1) after decays-in-flight of pions
($\pi$DIF), and  decays
coming from previously stopped (``old'') muons remaining in the target area (L3):

\noindent
\begin{tabular}{ll}
L1: $F_{\rm L1} = \frac{\lambda_{\pi} \lambda_{\mu}}{\lambda_{\pi}-\lambda_{\mu}}
(e^{-\lambda_{\mu}t} - e^{-\lambda_{\pi}t})$
\hspace*{0.5cm} & for $t> 0$,\\
L2: $F_{\rm L2} = \lambda_{\mu} e^{-\lambda_{\mu}t} $
\hspace*{0.5cm} & for $t> 0$, and\\
L3: $F_{\rm L3} = \lambda_{\mu} e^{-\lambda_{\mu}t} $
\hspace*{0.5cm} & for any $t$.\\
\end{tabular}

\noindent
The distribution coming from the presence of plural 
muons in the target area was
estimated to be $<$0.01 \%, and was ignored in the fit. 
The low-energy fraction of $\pi^+ \rightarrow \mbox{e}^+ \nu$
events due to shower leakage and radiative decays
was also negligible in the low-energy time spectrum fit. 

%%\subsubsection{High-energy time spectrum}

The primary time distribution component in the high-energy region
was the $\pi^+ \rightarrow \mbox{e}^+ \nu$  decay 
(H1: $F_{\rm H1} = \lambda_{\pi} e^{-\lambda_{\pi}t}$ for $t > 0$).
The amplitude of H1 also included the high-energy 
portion (E$_{e^+} \geq \rm{E}_{cut}$)
of decay-in-flight of muons ($\mu$DIF) following $\pi^+ \rightarrow \mu^+ \nu$ decay
at rest,
which was estimated by simulation \cite{geant}
to be $(2.07 \pm 0.06) \times 10^{-7}$ of L1.
The major backgrounds (H2) in the high-energy region came from 
muon decays due to the energy resolution of the detector,
radiative muon decays
in which the $\gamma$-ray raised the observed calorimeter energy
above E$_{cut}$, and 
extra hits (pile-up) in the calorimeter with a flat time distribution 
($e.g.$ due to neutrons from the pion production target). 
The H2 component had an identical time dependence to the low-energy 
spectrum (L1+L2).
The contribution from L3 via the same mechanism
was separately
treated as a muon decay component (H3)
to include other contributions of ``old'' muons.

Radiative pion decays $\pi^+ \rightarrow \mu^+ \nu \gamma$ 
(branching fraction, $2 \times 10^{-4}$ \cite{radpi})
followed by  $\mu^+ \rightarrow \mbox{e}^+ \nu \overline{\nu}$ decays 
could contribute to the high-energy region if the $\gamma$-ray
hit the calorimeter. The contribution of the extra $\gamma$-ray to 
the observed positron energy 
varied with the time difference of the two decays.
This contribution (H4) was simulated using the observed
pulse shapes  of the NaI(T$\ell$) and CsI signals, and is shown by the dashed line in Fig. \ref{time}b. 
The amplitude of this component was $(4.9 \pm 1.0) \times 10^{-7}$
of L1 in the fit.

The background in the region
$t < 0$ ns
was due to events with time distribution H3 and
 those in which a positron from an ``old'' muon hit T1
in coincidence with
a positron, from $\pi^+ \rightarrow \mu^+ \rightarrow \mbox{e}^+$ decay 
of the stopped pion, that missed T1
but hit the calorimeter, raising the observed energy above E$_{cut}$.
The shape of this time spectrum (H5),
including the inverse combination, was generated by simulation 
using the observed pulse shapes and the
energy distributions for the corresponding event topologies.
The shape and the relative amplitude of H5 are shown 
by the dotted line in Fig. \ref{time}b.

A pile-up cut based on the T1 waveform rejected events with two hits.
However, events with two T1 hits within the double pulse
resolution of T1 ($\Delta {\rm T} = 15.7 \pm 0.3$ ns) were accepted, and
the probability for the measured positron energy to be E$_{e^+} \geq$ E$_{cut}$
was high in those events.
By artificially increasing the double pulse resolution up to 200 ns,
the amplitude of this component (H6) was obtained and
fixed to the ``old'' muon background L3 in the fit.
The H6 component is shown by the full line in Fig. \ref{time}b;
the uncertainty in $R_{e/\mu}$ was 0.01 \%.

%%\subsubsection{Fit}

The free parameters in the fit for the low-energy region were
the amplitudes of L1, L2, and L3.
The time origin $t_0$,
which was determined using prompt events, was fixed in the fit.
The choice of $t_0$
did not affect the branching ratio as long as the amplitude of
L2 was a free parameter.
The free parameters for the high-energy region were
the amplitudes of
H1, H2, H3 and H5. 
The total $\chi^2$ of the high-energy and low-energy  fit was minimized with 
a common $t_0$. The fitting region was 
--290 ns to 520 ns 
excluding the prompt region of
--19 to 4 ns.

The overall fit result is almost indistinguishable from the data
and is not displayed in Fig. \ref{time}, except in (b) for $t<0$ ns
(thick solid line).
No structure was evident in the plot of residuals of the fit.
The raw branching ratio and its statistical and systematic uncertainties were 
 $R_{e/\mu}^{Raw} = (1.1972 \pm 0.0022(stat) \pm 0.0005(syst)) \times 10^{-4}$
with $\chi^2 / {\rm DOF} = 1.02$ (DOF=673).
The systematic uncertainty includes uncertainties of 
the parameters and shapes in the fit
and of small components excluded from the standard fitting function
as listed in Table \ref{result}.
 The branching ratio was stable for the fits with free pion and muon lifetimes,
which were consistent with the current values \cite{pdg}.

%\subsection{Corrections}

%\subsubsection{Monte Carlo}

Some corrections applied to the raw branching ratio
relied on simulation \cite{geant}.
Pions were generated 0.5 m upstream of the detector 
according to the measured pion beam distribution.
Small energy-dependent effects in the energy-loss processes 
of positrons change the relative acceptances
of low- and high-energy events. 
The ratio of the acceptances of the 
$\pi^+ \rightarrow \mu^+ \rightarrow \mbox{e}^+$ and 
$\pi^+ \rightarrow \mbox{e}^+ \nu$ decays was found
to be $0.9991 \pm 0.0003 (syst)$
for a WC3 radius cut  $r \leq$60 mm.

%\subsubsection{Tail correction}

The largest correction to the raw branching ratio 
was for the $\pi^+ \rightarrow \mbox{e}^+ \nu$ events below E$_{cut}$,
which primarily arose from the response function of the calorimeter.
Because of structure in the response function  
due to hadronic interactions \cite{bump}, 
which was not well reproduced by the simulation,
empirical measurements were performed.
Special data, using a simplified setup consisting of T2 and WC1-3,
taken with a 70-MeV/$c$ positron beam 
at various entrance angles were used 
to determine the response function.
In order to obtain the fraction of the 
$\pi^+ \rightarrow \mbox{e}^+ \nu$ events below E$_{cut}$
for the full setup, the difference in the detector geometry,
the $\pi^+ \rightarrow \mbox{e}^+ \nu$ angular distributions, and
radiative pion decays were estimated using simulation.
The fraction of
the events below E$_{cut} = 52$ MeV was found
to be $3.19 \pm 0.03(stat) \pm 0.08 (syst)$ \%.
Since a small contribution to the observed fraction from
low-energy positrons in the beam could
not be ruled out, the tail correction obtained in this way 
was treated as an upper bound.

In order to estimate the lower bound
to the tail fraction, 
$\pi^+ \rightarrow \mu^+ \rightarrow \mbox{e}^+$ events were suppressed
using an early decay-time region 5--35 ns,
pulse shape and total energy in B3, 
and measurements of the straightness of the pion track
\cite{neutrino}.
The resulting background-suppressed positron energy spectrum 
is shown by the shaded histogram in Fig. \ref{energy}.
The remaining background was subtracted from the spectrum
using the fact that
the background-suppressed spectrum in a low-energy region
contained a negligible $\pi^+ \rightarrow \mbox{e}^+ \nu$ tail contribution.
The area of the low-energy region was scaled 
to the full region ($< {\rm E}_{cut}$) using the 
known background distribution. This resulted in a lower bound 
of $1.48 \pm 0.07 (stat) \pm 0.08 (syst)$ \%.
Since the total energy cut used in the suppression 
method tended to remove $\pi^+ \rightarrow \mbox{e}^+ \nu$ events
with Bhabha scattering which resulted in larger energy deposit in B3,
a correction of $1.48 \pm 
 0.02 (syst)$ \% 
obtained by simulation was added
to the tail correction. Thus, the lower bound 
was  $2.95 \pm 0.07 (stat) \pm 0.08 (syst)$ \%.
Combining the upper and lower bounds,
a multiplicative tail correction of 
$1.0316 \pm 0.0012$ was obtained.

%\subsubsection{Other corrections}

Possible energy-dependent effects on $t_0$ were 
studied using positrons in the beam
at momenta 10--70 MeV/$c$,
and with positrons from muons stopped at the center of B3 
by lowering the beam momentum to
62 MeV/$c$. The multiplicative correction from this effect
was $1.0004 \pm 0.0005$.
Other uncertainties included are for possible trigger inefficiencies
($\pm 0.0003$) and distortions 
due to pile-up and other cuts ($\pm 0.0005$).

%\subsection{Systematic studies}

\begin{figure}[htb]
\centering
\includegraphics*[width=8.3cm]{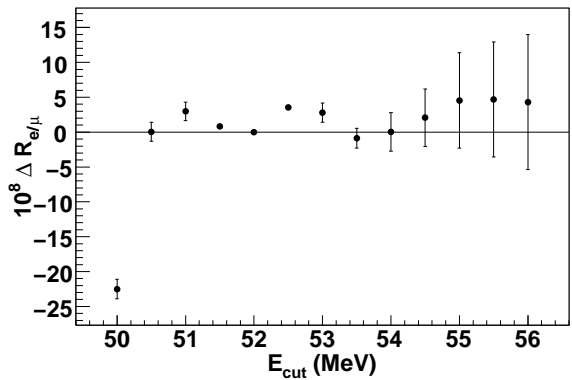}
\caption{Dependence of the branching ratio on E$_{cut}$
with respect to the value at 52 MeV.
The error bars indicate additional statistical and systematic uncertainties.
The variations indicated here are small compared to
the statistical uncertainty
of $23 \times 10^{-8}$ at E$_{cut} = 52$ MeV.
}
\label{ecut}
\end{figure}

Stability of the measured branching ratio was further tested 
for dependence on many parameters, 
such as 
fitting ranges, fiducial cuts, pile-up cuts and 
E$_{cut}$,
which provided confidence 
in the validity of the background functions and corrections.
Figure \ref{ecut} shows the dependence on E$_{cut}$.
The drop below 50.5 MeV is primarily due to the energy 
threshold of the HE-trigger.

%\section{Results and conclusion}

Table \ref{result} shows a summary of the fit uncertainties 
and corrections after ``unblinding''.
The measured branching ratio is 
$R_{e/\mu}^{Exp} = (1.2344 \pm 0.0023 (stat) \pm 0.0019 (syst)) \times 10^{-4}$,
consistent with previous work
and the SM prediction. 
The present result improves the test of e-$\mu$ universality
compared to previous experiments
by a factor of two:
$g_e / g_{\mu} = 0.9996 \pm 0.0012$ for the charged current.
Results using 
an order of magnitude more
data and possibly improved systematic uncertainty estimates
will be forthcoming.

This measurement also results in improved 90 \% confidence-level
limits \cite{fc} on the neutrino mixing parameter
$U_{ei}$ between the weak electron-neutrino eigenstate and 
a hypothetical mass eigenstate
$m_{\nu_i}$ \cite{neutrino},
$|U_{ei}|^2 < 0.0033 / (\rho_e -1)$
in the mass region $< 55$ MeV,
where $\rho_e$ is a kinetic factor found in Refs. \cite{shrock,oldnu}.

\begin{table}[thb]
\centering
\vspace*{0.5cm}
\begin{tabular}{l|rrr}
\hline
\hline
 & Values & \multicolumn{2}{c}{Uncertainties} \\
 & &\hspace*{0.5cm} $Stat$ &\hspace*{0.5cm} $Syst$\\
\hline
$R_{e/\mu}^{Raw}$ $(10^{-4})$ & 1.1972& 0.0022&  0.0005\\
\hspace{0.5cm} $\pi$,$\mu$ lifetimes & & & 0.0001\\
\hspace{0.5 cm} Other parameters & & & 0.0003\\
\hspace{0.5 cm} Excluded components & & & 0.0005\\
\hline
Corrections\\
\hspace{0.5 cm} Acceptance & 0.9991 & & 0.0003\\
\hspace{0.5 cm} Low-energy tail & 1.0316& & 0.0012\\
\hspace{0.5 cm} Other & 1.0004& & 0.0008\\
\hline
$R_{e/\mu}^{Exp}$ $(10^{-4})$ & 1.2344& 0.0023& 0.0019\\
\hline
\hline
\end{tabular}
\caption{The table includes the raw branching ratio with its 
statistical and systematic uncertainties, the multiplicative 
corrections with their errors, and the result after applying 
corrections.
}
\label{result}
\end{table}

%\section{Acknowledgment}

This work was supported
 by the Natural Sciences and Engineering Research Council
and TRIUMF through a contribution from the National Research Council of Canada,
and by the Research Fund for the Doctoral Program of Higher Education of China, 
by CONACYT doctoral fellowship from Mexico, and
by JSPS KAKENHI Grant numbers 18540274, 21340059, 24224006 in Japan.
We are grateful to Brookhaven National Laboratory for
the loan of the crystals, and to the TRIUMF operations, detector, 
electronics and DAQ groups for
their engineering and technical support.
\\

\end{document}